\documentclass[reprint,superscriptaddress,amsmath,amssymb,aps]{revtex4-1}

\usepackage{graphicx}
\usepackage{dcolumn}
\usepackage{bm}
\usepackage{hyperref}
\usepackage{xcolor}
\usepackage{tikz}

\definecolor{lime}{HTML}{A6CE39}
\DeclareRobustCommand{\orcidicon}{
	\begin{tikzpicture}
	\draw[lime, fill=lime] (0,0) 
	circle [radius=0.16] 
	node[white] {{\fontfamily{qag}\selectfont \tiny ID}};
	\draw[white, fill=white] (-0.0625,0.095) 
	circle [radius=0.007];
	\end{tikzpicture}
	\hspace{-2mm}
}
\foreach \x in {A, ..., Z}{\expandafter\xdef\csname orcid\x\endcsname{\noexpand\href{https://orcid.org/\csname orcidauthor\x\endcsname}
			{\noexpand\orcidicon}}
}

\newcolumntype{C}[1]{>{\centering\arraybackslash}p{#1}}

\usepackage{lineno}

\begin{document}

\title{Ruling out color transparency in quasi-elastic $^{12}$C(e,e$^{'}$p) up to $Q^2$ of 14.2 (GeV/c)$^2$}
\newcommand*{\MSU }{Mississippi State University, Mississippi State, Mississippi 39762, USA}
\newcommand*{\MSUindex}{1}
\affiliation{\MSU}
\newcommand*{\UVA }{University of Virginia, Charlottesville, Virginia 22903, USA}
\newcommand*{\UVAindex}{2}
\affiliation{\UVA}
\newcommand*{\JLAB }{Thomas Jefferson National Accelerator Facility, Newport News, Virginia 23606, USA}
\newcommand*{\JLABindex}{3}
\affiliation{\JLAB}
\newcommand*{\REG }{University of Regina, Regina, Saskatchewan S4S 0A2, Canada}
\newcommand*{\REGindex}{4}
\affiliation{\REG}
\newcommand*{\NCAT }{North Carolina A \& T State University, Greensboro, North Carolina 27411, USA}
\newcommand*{\NCATindex}{5}
\affiliation{\NCAT}
\newcommand*{\KENT }{Kent State University, Kent, Ohio 44240, USA}
\newcommand*{\KENTindex}{6}
\affiliation{\KENT}
\newcommand*{\ZAG }{University of Zagreb, Zagreb, Croatia}
\newcommand*{\ZAGindex}{7}
\affiliation{\ZAG}
\newcommand*{\TEMP }{Temple University, Philadelphia, Pennsylvania 19122, USA}
\newcommand*{\TEMPindex}{8}
\affiliation{\TEMP}
\newcommand*{\YER }{A.I. Alikhanyan  National  Science  Laboratory \\ (Yerevan  Physics
Institute),  Yerevan  0036,  Armenia}
\newcommand*{\YERindex}{9}
\affiliation{\YER}
\newcommand*{\WM }{The College of William \& Mary, Williamsburg, Virginia 23185, USA}
\newcommand*{\WMindex}{10}
\affiliation{\WM}
\newcommand*{\CUA }{Catholic University of America, Washington, DC 20064, USA}
\newcommand*{\CUAindex}{11}
\affiliation{\CUA}
\newcommand*{\HU }{Hampton University, Hampton, Virginia 23669, USA}
\newcommand*{\HUindex}{12}
\affiliation{\HU}
\newcommand*{\FIU }{Florida International University, University Park, Florida 33199, USA}
\newcommand*{\FIUindex}{13}
\affiliation{\FIU}
\newcommand*{\CNU }{Christopher Newport University, Newport News, Virginia 23606, USA}
\newcommand*{\CNUindex}{14}
\affiliation{\CNU}
\newcommand*{\JAZ }{Jazan University, Jazan 45142, Saudi Arabia}
\newcommand*{\JAZindex}{15}
\affiliation{\JAZ}
\newcommand*{\UTENN }{University of Tennessee, Knoxville, Tennessee 37996, USA}
\newcommand*{\UTENNindex}{16}
\affiliation{\UTENN}
\newcommand*{\OHIO }{Ohio University, Athens, Ohio 45701, USA}
\newcommand*{\OHIOindex}{17}
\affiliation{\OHIO}
\newcommand*{\UCONN }{University of Connecticut, Storrs, Connecticut 06269, USA}
\newcommand*{\UCONNindex}{18}
\affiliation{\UCONN}
\newcommand*{\SBU }{Stony Brook University, Stony Brook, New York 11794, USA}
\newcommand*{\SBUindex}{19}
\affiliation{\SBU}
\newcommand*{\ODU }{Old Dominion University, Norfolk, Virginia 23529, USA}
\newcommand*{\ODUindex}{20}
\affiliation{\ODU}
\newcommand*{\ANL }{Argonne National Laboratory, Lemont, Illinois 60439, USA}
\newcommand*{\ANLindex}{21}
\affiliation{\ANL}
\newcommand*{\BOULDER }{University of Colorado Boulder, Boulder, Colorado 80309, USA}
\newcommand*{\BOULDERindex}{22}
\affiliation{\BOULDER}
\newcommand*{\ORSAY }{Institut de Physique Nucleaire, Orsay, France}
\newcommand*{\ORSAYindex}{23}
\affiliation{\ORSAY}
\newcommand*{\UNH }{University of New Hampshire, Durham, New Hampshire 03824, USA}
\newcommand*{\UNHindex}{24}
\affiliation{\UNH}
\newcommand*{\JMU }{James Madison University, Harrisonburg, Virginia 22807, USA}
\newcommand*{\JMUindex}{25}
\affiliation{\JMU}
\newcommand*{\RUTG }{Rutgers University, New Brunswick, New Jersey 08854, USA}
\newcommand*{\RUTGindex}{26}
\affiliation{\RUTG}
\newcommand*{\CMU }{Carnegie Mellon University, Pittsburgh, Pennsylvania 15213, USA}
\newcommand*{\CMUindex}{27}
\affiliation{\CMU}

\author{D. Bhetuwal}\affiliation{\MSU}
\author{J. Matter}\affiliation{\UVA}
\author{H. Szumila-Vance}\affiliation{\JLAB}
\author{M.~L.~Kabir}\affiliation{\MSU}
\author{D. Dutta}\affiliation{\MSU}
\author{R.~Ent}\affiliation{\JLAB}  

\author{D.~Abrams}\affiliation{\UVA} 
\author{Z.~Ahmed}\affiliation{\REG}  
\author{B.~Aljawrneh}\affiliation{\NCAT}  
\author{S.~Alsalmi}\affiliation{\KENT} 
\author{R.~Ambrose}\affiliation{\REG} 
\author{D.~Androic}\affiliation{\ZAG}  
\author{W.~Armstrong}\affiliation{\TEMP} 
\author{A.~Asaturyan}\affiliation{\YER} 
\author{K.~Assumin-Gyimah}\affiliation{\MSU}   
\author{C.~Ayerbe Gayoso}\affiliation{\WM}\affiliation{\MSU}    
\author{A.~Bandari}\affiliation{\WM}   
\author{S.~Basnet} \affiliation{\REG}  
\author{V.~Berdnikov}\affiliation{\CUA}    
\author{H.~Bhatt}\affiliation{\MSU}   
\author{D.~Biswas}\affiliation{\HU}   
\author{W.~U.~Boeglin}\affiliation{\FIU} 
\author{P.~Bosted}\affiliation{\WM} 
\author{E.~Brash}\affiliation{\CNU}    
\author{M.~H.~S.~Bukhari}\affiliation{\JAZ}  
\author{H.~Chen}\affiliation{\UVA}             
\author{J.~P.~Chen}\affiliation{\JLAB}           
\author{M.~Chen}\affiliation{\UVA}             
\author{E.~M.~Christy}\affiliation{\HU}          
\author{S.~Covrig}\affiliation{\JLAB}           
\author{K.~Craycraft}\affiliation{\UTENN}        
\author{S.~Danagoulian}\affiliation{\NCAT}     
\author{D.~Day}\affiliation{\UVA}             
\author{M.~Diefenthaler}\affiliation{\JLAB}     
\author{M.~Dlamini}\affiliation{\OHIO}          
\author{J.~Dunne}\affiliation{\MSU}            
\author{B.~Duran}\affiliation{\TEMP}            
\author{R.~Evans}\affiliation{\REG}            
\author{H.~Fenker}\affiliation{\JLAB}           
\author{N.~Fomin}\affiliation{\UTENN}            
\author{E.~Fuchey}\affiliation{\UCONN}           
\author{D.~Gaskell}\affiliation{\JLAB}          
\author{T.~N.~Gautam}\affiliation{\HU}         
\author{F.~A.~Gonzalez}\affiliation{\SBU}       
\author{J.~O.~Hansen}\affiliation{\JLAB}           
\author{F.~Hauenstein}\affiliation{\ODU}       
\author{A.~V.~Hernandez}\affiliation{\CUA}       
\author{T.~Horn}\affiliation{\CUA}             
\author{G.~M.~Huber\orcidA{}}\affiliation{\REG}       
\author{M.~K.~Jones}\affiliation{\JLAB}          
\author{S.~Joosten}\affiliation{\ANL}          
\author{A.~Karki}\affiliation{\MSU}            
\author{C.~Keppel}\affiliation{\JLAB}           
\author{A.~Khanal}\affiliation{\FIU}           
\author{P.~M.~King}\affiliation{\OHIO}             
\author{E.~Kinney}\affiliation{\BOULDER}           
\author{H.~S.~Ko}\affiliation{\ORSAY}            
\author{M.~Kohl}\affiliation{\HU}              
\author{N.~Lashley-Colthirst}\affiliation{\HU}        
\author{S.~Li}\affiliation{\UNH}               
\author{W.~B.~Li}\affiliation{\WM}               
\author{A.~H.~Liyanage}\affiliation{\HU}         
\author{D.~Mack}\affiliation{\JLAB}              
\author{S.~Malace}\affiliation{\JLAB}           
\author{P.~Markowitz}\affiliation{\FIU}       
\author{D.~Meekins}\affiliation{\JLAB}          
\author{R.~Michaels}\affiliation{\JLAB}         
\author{A.~Mkrtchyan}\affiliation{\YER}         
\author{H.~Mkrtchyan}\affiliation{\YER}        
\author{S.J.~Nazeer}\affiliation{\HU}         
\author{S.~Nanda}\affiliation{\MSU}   
\author{G.~Niculescu}\affiliation{\JMU}        
\author{I.~Niculescu}\affiliation{\JMU}        
\author{D.~Nguyen}\affiliation{\UVA} 
\author{Nuruzzaman}\affiliation{\RUTG}       
\author{B.~Pandey}\affiliation{\HU}           
\author{S.~Park}\affiliation{\SBU}             
\author{E.~Pooser}\affiliation{\JLAB}           
\author{A.~Puckett}\affiliation{\UCONN}         
\author{M.~Rehfuss}\affiliation{\TEMP}          
\author{J.~Reinhold}\affiliation{\FIU}         
\author{N.~Santiesteban}\affiliation{\UNH}     
\author{B.~Sawatzky}\affiliation{\JLAB}          
\author{G.~R.~Smith}\affiliation{\JLAB}             
\author{A.~Sun}\affiliation{\CMU}               
\author{V.~Tadevosyan}\affiliation{\YER}        
\author{R.~Trotta}\affiliation{\CUA}           
\author{S.~A.~Wood\orcidB{}}\affiliation{\JLAB}            
\author{C.~Yero} \affiliation{\FIU}  
\author{J.~Zhang}\affiliation{\SBU}        
\collaboration{for the Hall C Collaboration}
\noaffiliation

\date{\today}

\begin{abstract}
Quasielastic $^{12}$C$(e,e'p)$ scattering was measured at space-like 4-momentum transfer squared $Q^2$~=~8, 9.4, 11.4, and 14.2 (GeV/c)$^2$, the highest ever achieved to date. Nuclear transparency for this reaction was extracted by comparing the measured yield to that expected from a plane-wave impulse approximation calculation without any final state interactions. The measured transparency was consistent with no $Q^2$ dependence, up to proton momenta of 8.5~GeV/c, ruling out the quantum chromodynamics effect of color transparency at the measured $Q^2$ scales in exclusive $(e,e'p)$ reactions. These results impose strict constraints on models of color transparency for protons.  
\end{abstract}

\maketitle 


At low energies, the strong interaction is well described in terms of nucleons (protons and neutrons) exchanging mesons~\cite{gfmcrmp15}, whereas at high energies, perturbative Quantum Chromodynamics (pQCD) characterizes the strong force in terms of quarks and gluons carrying color charge. Although these two descriptions are well understood in their respective energy scales, the transition between them is not uniquely identified. Quantum Chromodynamics (QCD) predicts that protons produced in exclusive processes at sufficiently high 4-momentum transfer ($Q$), will experience suppressed final (initial) state interactions resulting in a significant enhancement in the nuclear transparency (T)~\cite{CT, CT_brodsky}. This unique prediction of QCD is named color transparency (CT), and the observation of the onset of CT may help identify the transition between the two alternate descriptions of the strong force. \\

Mueller and Brodsky~\cite{CT, CT_brodsky} introduced CT as a direct consequence of the concept that in exclusive processes at sufficiently high momentum transfer, hadrons are produced in a point-like configuration (PLC). Quantum mechanics accounts for the existence of hadrons that fluctuate to a PLC, and a high momentum transfer virtual photon preferentially interacts with a hadron in a PLC (with transverse size $r_{\perp}\approx 1/Q$)~\cite{factor1}. The reduced transverse size, color neutral PLC is screened from external fields, analogous to
a reduced transverse size electric dipole~~\cite{factor1}. At sufficiently high Lorentz factor, the PLC maintains its compact size long enough to traverse the nuclear volume while experiencing reduced interaction with the spectator nucleons. It thereby experiences reduced attenuation in the nucleus due to color screening and the properties of the strong force~\cite{factor1}. The onset of CT is thus a signature of QCD degrees of freedom in nuclei and is expected to manifest as an increase in $T$ with increasing momentum transfer.\\

The energy regime for the onset of CT is not precisely known but provides crucial insights for nuclear theory, see a summary in Ref~\cite{ctreview2012}. The suppression of further interactions with the nuclear medium is a fundamental assumption necessary to account for Bjorken scaling in deep-inelastic scattering at small $x_B$~\cite{FS88}. Moreover, the onset of CT is of specific interest as it can help identify the relevant space-like 4-momentum transferred squared ($Q^2$) where factorization theorems are applicable~\cite{strikman} enabling the extraction of Generalized Parton Distributions (GPDs)~\cite{gpd1,gpd2}. At intermediate energies, there exists a trade-off between the selection of the PLC and its expansion as it transits the nucleus. Therefore, the onset of CT is best observed at the intermediate energy regime where the expansion distance of the PLC becomes significant compared to the nuclear radius. Theory anticipates that it is more probable to observe the onset of CT at lower energies for meson production than for baryons as it is more probable for quark-antiquark pairs (mesons) to form a PLC than three quark systems (baryons)~\cite{eides}. Additionally, the significantly larger Lorentz factor for mesons ensures that the expansion distance over which the PLC evolves back to its equilibrium configuration can be as large as the nuclear radius at lower energies for mesons than for baryons~\cite{vgg}.\\

The predicted onset of CT for final-state mesons has been demonstrated in several experiments at Jefferson Lab (JLab). Pion photoproduction cross sections of $^4$He to $^2$H were found to be consistent with CT theories showing a positive rise in the ratio~\cite{he4tranprc}. Precise and systematic studies of pion electroproduction on a range of targets established a positive slope in the transparency ratios for $Q^2$ in the range from 1--5~(GeV/c)$^2$, as well as an $A$-dependence of the slope. These results were found to be consistent with models that include CT~\cite{ben_prl,xqian}. The onset of CT in mesons was further confirmed by a JLab experiment measuring the nuclear transparency of $\rho^{0}$ electroproduction which showed slopes vs $Q^2$ consistent with the same CT models~\cite{elfassi12} as the pions. While empirical evidence conclusively confirms the onset of CT in mesons at momentum scales corresponding to $Q^2 \approx$~1 (GeV/c)$^2$, the observation of the onset of CT in baryons is somewhat ambiguous.\\ 

In a pioneering experiment at the Brookhaven National Lab (BNL), the E850 collaboration attempted to measure the onset of CT using the proton knockout $A(p,2p)$ reaction at $\approx$ 90$^{\circ}$ c.m. angle~\cite{bnl}. The nuclear transparency was measured as the ratio of the quasielastic cross section from a nuclear target to that of the free $pp$ cross section which varies rapidly with $Q^2$~\cite{ralpire,ralston}. The transparency was measured as a function of an effective beam momentum, $P_{\text{eff}}$, and was shown to have a positive rise from $P_{\text{eff}} =\,$5.9--9~GeV/c~\cite{bnl}. A subsequent decrease in the transparency was observed between $P_{\text{eff}} =$ 9.5--14.4~GeV/c~\cite{bnlnew1,bnlnew2,bnlfinal}. This enhancement and subsequent fall in the nuclear transparency spans a $Q^2$ (Mandelstam $-t$) range of 4.8--12.7 (GeV/c)$^2$ and outgoing proton momentum range of 3.3--7.7 GeV/c. Two possible explanations for the decrease in transparency at the higher momenta are: an in-medium suppression of the energy dependence of the $pp$ elastic cross section known as nuclear filtering~\cite{ralpire,ralston}, or the excitation of charmed quark resonances or other exotic multi-quark states~\cite{brodsky_ch}.\\


Hadron propagation through the nuclear medium is dominated by a reduction of flux at high energies. In the $A(p,2p)$ reaction both the incoming and outgoing protons experience a reduction of flux making it more challenging to interpret. Subsequently, the ambiguous results from the BNL experiment were investigated with the $(e,e'p)$ process, which employs electrons, a weakly interacting probe, to avoid the complication of the reduction of flux of the hadronic probe.
In quasielastic scattering of electrons from a nucleus, $A(e,e'p)$, the outgoing proton can interact with the spectator nucleons such that absorption and rescattering of the outgoing proton results in a reduction of the measured $A(e,e'p)$ yield. 
Furthermore, compared to the $(p,2p)$ process, the elementary elastic $ep$ scattering cross section is accurately known and smoothly varying with energy transfer, and the $A(e,e'p)$ process is less sensitive to the poorly known large momentum components of the nuclear wave function~\cite{farrar89}. The underlying processes that contribute to the $A(e,e'p)$ yield such as multiple scattering, and initial/final state interactions are energy independent. Thus, in the nucleon-meson picture of the nucleus, one would expect that the transparency, $T$, defined as the ratio of the measured $A(e,e'p)$ yield to that calculated in the Plane Wave Impulse Approximation, should also be independent of energy. Measurement of $T$ can therefore test for deviation from the expectations of conventional nuclear physics and the onset of quark-gluon degrees of freedom.\\

Previous $A(e,e'p)$ experiments~\cite{ne18_1,ne18_2,donprl,garrow} have measured the nuclear transparency of protons on a variety of nuclei up to $Q^2 =$ 8.1 (GeV/c)$^2$. These experiments yielded missing energy and momentum distributions consistent with conventional nuclear physics and did not observe any significant $Q^2$ dependence in the nuclear transparency.
This ruled out the onset of CT for protons at $Q^2$ values corresponding to outgoing proton momenta of 5 GeV/c, which in some interpretations is just before the rise of transparency noted in the $A(p,2p)$ data.\\

The recent 12~GeV upgrade at JLab allows access to the entire $Q^2$ range and outgoing proton momentum range of the BNL experiment for the first time. It also allows significant overlap between the knocked out proton momentum in electron scattering and the effective proton momentum quoted by the BNL $A(p,2p)$ experiment, within the range where the enhancement in nuclear transparency was observed~\cite{bnl}. These features make it possible to explore all possible independent variables ($Q^2$, incident or outgoing proton momentum) that could be driving the enhancement in transparency observed in the BNL experiment. In this letter, we report on the latest quasi-elastic electron scattering experiment to search for the onset of CT at the upgraded JLab. This experiment extends the nuclear transparency measurements in $^{12}C(e,e'p)$ to the highest $Q^2$ to date and covers the complete kinematic phase space of the enhancement observed by the BNL experiment.\\ 

The experiment was carried out in Hall C at JLab
and used the continuous wave electron beam with beam energies of 6.4 and 10.6~GeV and beam currents up to 65~$\mu$A.
The total accumulated beam charge was determined with $\approx 1$\% uncertainty by a set of resonant-cavity based beam-current monitors and a parametric transformer monitor. The beam energy was determined with an uncertainty of 0.1\% by measuring the bend angle of the beam, on its way into Hall C, as it traversed a set of magnets with precisely known field integrals. The main production target was a carbon foil of 4.9\% radiation lengths (rl), while a second carbon foil of 1.5\% rl was used for systematic studies. The thickness of the foils was measured to better than 0.5\%. A 10-cm-long (726 mg/cm$^2$) liquid hydrogen target was used to measure the elementary $ep$ scattering process. Two aluminum foils placed 10-cm apart were used to monitor the background from the aluminum end caps of the hydrogen target cell. The measured $ep$ elastic cross section agrees with the world data~\cite{bosted95}, and a comparison to a Monte Carlo simulation~\cite{simc} yields an overall normalization uncertainty of 1.8\% (see the supplementary material).\\

The scattered electrons were detected in the legacy High Momentum Spectrometer (HMS, momentum acceptance $\Delta p/p\pm 10\%$, solid angle $\Omega=7$~msr)~\cite{donprl} in coincidence with the knocked-out protons detected in the new Super High Momentum Spectrometer (SHMS, momentum acceptance $\Delta p/p$ from -10 to +12\%, solid angle $\Omega=4$~msr)~\cite{shms_nim}. The SHMS central angle was chosen to detect protons along the electron three-momentum transfer, $\vec{q}$. 
The kinematics of the experiment are listed in the supplementary material.\\

The solid angle of the spectrometers was defined for electrons and the coincident $(e,e'p)$ process by a 2-in-thick tungsten alloy collimator. The detector packages in the two spectrometers were similar, and they included four planes of segmented scintillators (except for the last plane in the SHMS which used quartz bars) that were used to form the trigger and to provide time-of-flight information. Two 6-plane drift chambers were used to measure particle tracks with better than 250~$\mu$m resolution. The tracking efficiency was continuously monitored with an uncertainty of $\approx$~0.1\% for the HMS and $<0.5\%$ for the SHMS. The uncertainty was obtained from the average variation of the tracking efficiency when using three independent criteria for determining the efficiency. The typical rms resolutions in the HMS (SHMS) were 0.2\% (0.1\%) for momentum, 0.8 (0.9) mrad for horizontal angle and 1.2 (1.1) mrad for the vertical angle. In the HMS, a threshold gas Cherenkov detector and a segmented Pb-glass calorimeter were used for electron identification. The protons in the SHMS were identified 
by coincidence time after excluding pions using a noble-gas threshold Cherenkov detector and a segmented Pb-glass calorimeter. The pion-to-electron ratio in the HMS ranged from $\approx$~10$^{-1}$ to 10$^{-3}$, while the pion-to-proton ratio in the SHMS was always $<0.2$. The corrections for particle energy loss through the spectrometers were determined to better than 1$\%$. The electron-proton coincidence events were recorded in 1-hour-long runs via a data acquisition system operated using the CEBAF Online Data Acquisition (CODA) software package~\cite{coda}. Prescaled singles (inclusive) electron and proton events were simultaneously recorded for systematic studies.
The coincidence time was determined as the difference in the time of flight between the two spectrometers with corrections to account for path-length variations from the central trajectory and the individual start-times. The coincidence time rms resolution was 380~ps, more than sufficient to resolve the individual bursts of the 4~ns beam structure. The rate of accidental coincidences was $<0.2\%$.\\

The electron beam energy/momentum ($E_e/\vec{p}_e$) and the energy/momentum of the scattered electron ($E_{e'}/\vec{p}_{e'}$) measured by the HMS were used to determine $\vec{q} = \vec{p}_{e} - \vec{p}_{e'}$ and the energy transfer $\nu = E_e - E_{e'}$ for each coincidence event. The kinetic energy ($T_{p'}$) and momentum ($\vec{p}_{p'}$) of knocked out protons measured in the SHMS were used to determine the missing energy $E_m = \nu - T_{p'} - T_{A-1}$ and missing momentum $\vec{p}_{m} = \vec{p}_{p'} - \vec{q}$ for the coincidence event, where $T_{A-1}$ is the reconstructed kinetic energy of the $A-1$ recoiling nucleus. The experimental yield on the $^{12}$C target was obtained by integrating the charge-normalized coincidence events over a phase space defined by $E_m <$80 MeV and $|\vec{p}_m| <$300 MeV/c. These constraints eliminate inelastic contributions due to pion production while integrating over the majority of the single particle wave function. The experimental yield was corrected for all known inefficiencies of both spectrometers such as the detector efficiencies (97\%\,-\,99\%), trigger efficiency (98\%\,-\,99\%), tracking efficiencies (99\%\,-HMS and 94\%\,-\,99\%\,-SHMS), computer and electronic livetimes (94\%\,-\,99\%), and proton absorption in the SHMS ($\approx$\,8\%). The systematic uncertainty arising from the cut dependence of the experimental yield was determined by varying the cuts one at a time and recording the variation in yields for the different kinematic settings. The quadrature sum of the variation over all the different cuts was used as the event selection uncertainty ($\approx$~1.4\%). The uncertainty due to the livetime and the detector and trigger efficiencies was determined from a set of luminosity scans on a $^{12}$C target, performed in each spectrometer immediately before and after the experiment. The charge-normalized yield from these scans for each spectrometer was found to be independent of the beam current within statistical uncertainties, and the average variation in the normalized yield vs beam current was recorded as the systematic uncertainty (0.5\%). 
The uncertainty due to the charge measurement was estimated to be $\approx 1$\% which was validated by the change in the charge-normalized experimental yield when varying the minimum beam current cut.\\
\\
A Monte Carlo simulation~\cite{simc} of the $A(e,e'p)$ process was performed assuming the plane-wave impulse approximation (PWIA) to be valid, in which case the $\vec{p}_m$ is equal to the
initial momentum of the proton in the carbon nucleus, and the cross section is calculated in a factorized form as:
\begin{equation}
\frac{d^6\sigma}{dE_{e'}d\Omega_{e'}dE_{p'}d\Omega_{p'}} = E_{p'}|p_{p'}|\sigma_{ep}S(E_m,\vec{p}_m),
\end{equation}
where $\Omega_{e'}$, $\Omega_{p'}$ are the solid angles of the outgoing electron and proton respectively, $\sigma_{ep}$ is the off-shell $ep$ cross section and $S(E_m,\vec{p}_m)$ is the spectral function defined as the joint probability of finding a proton with momentum $p_m$ and separation energy $E_m$ within the nucleus. The simulation used the De Forest $\sigma_{1}^{cc}$ prescription~\cite{deForest} for the off-shell cross section, and the simulated yield was insensitive ($<$ 0.1\%) to the off-shell effect. The independent particle spectral functions used in the simulation were the same as those used in Ref.~\cite{ne18_1,ne18_2,donprl,garrow}. The effect of nucleon-nucleon correlations, which cause the single particle strength to appear at high $E_m$, was included by applying a correction factor of 1.11 $\pm$ 0.03 as previously determined in Ref.~\cite{tomthesis}. The simulated yield was obtained by integrating over the same phase-space volume as for the experimental data. The total model-dependent uncertainty was 3.9\% when the uncertainty in the spectral function (2.8\%) and the corrections due to nucleon-nucleon correlations are combined in quadrature.\\
\begin{figure}[hbt!]
  \centering
\includegraphics[width=0.48\textwidth]{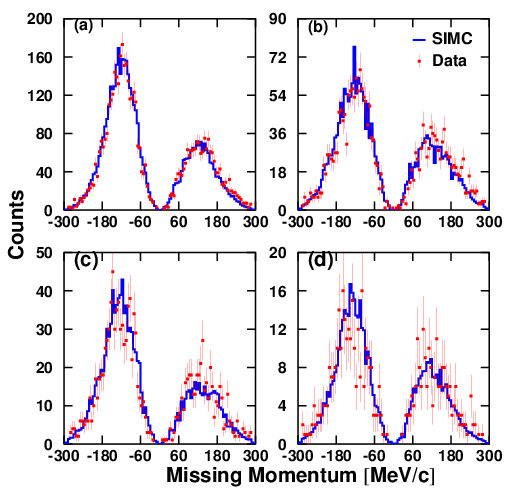}
 	 \caption[]{The missing momentum, $p_m$, for the carbon data is shown for each kinematic setting. (a) $Q^2 = 8.0$ (b) $Q^2 = 9.4$ (c) $Q^2 = 11.4$ and (d) $Q^2 = 14.2$\,(GeV/c)$^2$ }
 	 
  \label{fig:pmiss}
 \end{figure}

\begin{table}[htb!]
\caption{\label{errors} Systematic Uncertainties}
\centering
\begin{tabular}{lc}\hline 
Source & $Q^2$ dependent uncertainty (\%) \\ \hline
Spectrometer acceptance & 2.6\\
Event selection & 1.4 \\
Tracking efficiency & 0.5\\
Radiative corrections & 1.0 \\ 
Live time \& Det. efficiency  & 0.5 \\\hline
Source & Normalization uncertainty (\%) \\ \hline
Elastic $ep$ cross section & 1.8 \\
Target thickness & 0.5\\
Beam charge & 1.0\\
Proton absorption & 1.2\\\hline
Total & 4.0 \\ \hline
    \end{tabular}
\end{table} 
The measured $^{12}$C$(e,e'p)$ yields as a function of $p_m$ are shown in Fig.~\ref{fig:pmiss}, along with the simulated yields. The constraint of $E_m <$ 80 MeV was applied to both data and simulation. The shapes of the data and simulated distributions agree with each other very well for all four $Q^2$ settings, validating the use of the impulse approximation. It also indicates the robustness of the spectrometer models in the Monte Carlo simulation. The uncertainty from the spectrometer acceptance was estimated to be 2.6\% by comparing the measured and simulated focal plane positions and angles as well as the reconstructed angles and momenta at the reaction vertex. The $p_m$ distributions shown in Fig.~\ref{fig:pmiss} are very sensitive to the reconstructed momenta and angles and the average bin-by-bin difference between the data and simulated spectra normalized to each other was used as the systematic uncertainty due to acceptance.  
Table~\ref{errors} lists the major sources of systematic uncertainty. The total uncertainty is calculated as the quadrature sum. The model dependent uncertainty is not included in the table.\\ 
\begin{figure}[hbt!]
  \centering
\includegraphics[width=0.5\textwidth]{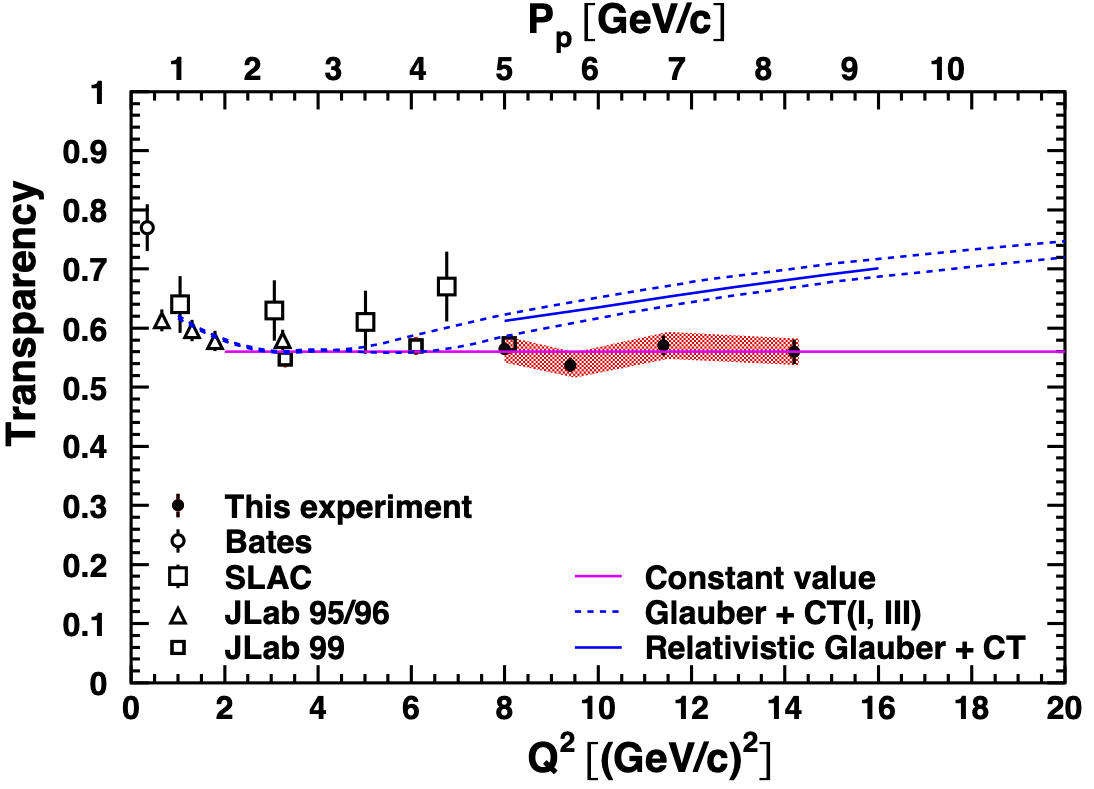}
 	 \caption[]{The carbon nuclear transparency from this experiment along with all previous experiments~\cite{batest,ne18_1,ne18_2,donprl,garrow}. The space-like 4-momentum transfer squared is shown along the $x-$axis (bottom scale), and the momentum of the knocked out proton is also shown along the top scale of the $x-$axis. The solid magenta line is for a constant value of 0.56. The dashed lines are theory predictions including CT~\cite{misak} for two different set of parameters and the solid blue line is a prediction from a relativistic Glauber calculation with CT~\cite{Cosyn}. The error bars show the statistical uncertainty while the band shows the 4.0\% systematic uncertainty. The 3.9\% model-dependent uncertainty is not shown.}
  \label{fig:C_all}
 \end{figure} 

The nuclear transparency was extracted as the ratio of experimental yield to the PWIA yield integrated over the same phase space volume $V$:
\begin{equation}
    T(Q^2) = \frac{\int_{V} d^3p_{m}dE_{m} Y_{\text{exp}}(E_m,\vec{p}_m)}{\int_{V} d^3p_{m}dE_{m} Y_{\text{PWIA}}(E_m,\vec{p}_m)},
\end{equation}
where $V$ is the phase space volume as defined earlier, $Y_{\text{exp}}(E_m,\vec{p}_m)$ is the experimental yield and $Y_{\text{PWIA}}(E_m,\vec{p}_m)$ is the PWIA yield.
The extracted nuclear transparency as a function of $Q^2$ is shown in Fig.~\ref{fig:C_all} along with all previous measurements. The model-dependent uncertainty is not shown in Fig.~\ref{fig:C_all} as to be consistent with the graphics of previous experiments. The measured nuclear transparency of carbon is found to be both energy and $Q^2$ independent up to $Q^2 =$ 14.2\,(GeV/c)$^2$, the highest accessed in quasi-elastic electron scattering to date. The combined data set from all measurements above $Q^2 =$ 3.0\,(GeV/c)$^2$ was fit to a constant value with a reduced $\chi^2$ of 1.3. The outgoing proton momentum of this experiment overlaps with the effective proton momentum of the BNL experiments that reported an enhancement in nuclear transparency~\cite{bnlfinal}. Moreover, the $Q^2$ and outgoing proton momentum of this experiment are significantly higher than the BNL experiment. As the underlying reaction mechanisms of the $A(p,2p)$ and $A(e,e'p)$ processes are different, these results provide key insight into the process dependence of exclusive scattering and the corresponding transparency. 
The differences governing the observed onset of CT for mesons at $Q^2$ 
of about 1 (GeV/c)$^2$ 
and the absence of the onset of CT for protons at more than an order-of-magnitude higher $Q^2$
may provide strong clues regarding the differences between two- and three-quark systems. Future experiments at JLab and elsewhere will further quantify such differences for pions, $\rho$-mesons and photons~\cite{e01107,kawtar,gluexprop}.\\


In summary, exclusive measurements were performed for $Q^2$ from 8--14.2~(GeV/c)$^2$ on hydrogen and carbon targets. The nuclear transparency extracted from these measurements is consistent with traditional nuclear physics calculations and does not support the onset of color transparency. The proton momentum scales accessed in this experiment rule out color transparency as the reason for a rise in transparency noted in the $A(p,2p)$ data. The present results probe down to a transverse-size as small as $\approx$~0.05 fm in the three-quark nucleon system, placing very strict constraints on the onset of color transparency at intermediate energies and all current models.\\
 
This work was funded in part by the U.S. Department of Energy, including contract 
AC05-06OR23177 under which Jefferson Science Associates, LLC operates Thomas Jefferson National Accelerator Facility, and by the U.S. National Science Foundation and the Natural Sciences and Engineering Research Council of Canada. We wish to thank the staff of Jefferson Lab for their vital support throughout the experiment. We are also grateful to all granting agencies providing funding support to authors throughout this project.

 
\bibliography{ctbibs}{}
\bibliographystyle{unsrt}


\end{document}